\documentclass[prl,twocolumn,aps,showpacs]{revtex4}
\usepackage{graphicx}
\usepackage{epsfig}
\usepackage{bm,hyperref}

\def\be{\begin{equation}}
\def\ee{\end{equation}}
\def\ba{\begin{eqnarray}}
\def\ea{\end{eqnarray}}

\begin{document}
\title
{Vector Potential and Berry phase-induced Force}
\author{Qi Zhang}
\affiliation{Institute of Physics, Chinese Academy of Sciences,
Beijing 100080, China}
\author{Biao Wu}
\email[Corresponding author:~]{bwu@aphy.iphy.ac.cn}
\affiliation{Institute of Physics, Chinese Academy of Sciences,
Beijing 100080, China}
\date{March 8th, 2006}
\begin{abstract}
We present a general theoretical framework for the exact treatment of 
a hybrid system that is composed of a quantum subsystem and a 
classical subsystem. When the quantum subsystem is dynamically fast and
the classical subsystem is slow, a vector potential is generated 
with a simple canonical transformation.
This vector potential, on one hand, gives rise to the familiar
Berry phase in the fast quantum dynamics; on the other hand, it
yields a Lorentz-like force in the slow classical dynamics.
In this way, the pure phase (Berry phase) of a wavefunction is linked 
to a physical force.
\end{abstract}
\pacs{03.65.Vf, 03.65.Ca}
\maketitle
Single spin detection was recently achieved with the magnetic-resonance-force 
microscopy\cite{Rugar2004Nature}. Technically, this remarkable
experiment marks a major step in human's ability to control objects at 
the atomic scale and has a great application potential in future 
technology\cite{Hammel2004Nature}. On the scientific side, this experiment 
also invokes many interesting questions of fundamental interest.

First, the system used in the experiment is a hybrid system composed of a 
quantum subsystem (single spin) and a classical subsystem (cantilever). 
In Ref.\cite{SlichterBook,Berman2002PRA}, both subsystems are treated classically. 
Although this is adequate to address specific issues and systems related 
to the current experiments\cite{Rugar2004Nature,Mamin2003PRL}, 
this method is inadequate to treat such a hybrid system in general,
for example, when the quantum subsystem is not a spin or when one
tries to discuss phase-related issues. This inadequacy will become
more transparent as our discussion proceeds. 
Then the question is, how to treat exactly a general hybrid system? This question
will grow more important since various techniques are being developed or explored
to control objects at the atomic scale. These techniques
are certainly all based on hybrid systems where a classical sensor
interacts with a quantum object.

Secondly, this experiment has another interesting aspect, 
that is, the quantum subsystem (single spin) is 
dynamically fast and the classical subsystem (cantilever) is slow. 
This is very similar to the systems treated in the Born-Oppenheimer 
approximation\cite{Mead1979JCP,Aharonov1990PRL}
and other similar contexts\cite{Iida1985PTP,Stone1986PRD}, 
where one subsystem is fast and the other is slow. 
In these ``Born-Oppenheimer'' systems, where both of the subsystems 
are quantal, vector potentials related to Berry phase\cite{Berry1984PRS}
were found to arise and generate physical consequences. Can a 
similar vector potential be found in hybrid systems? If so, 
what physical effects can it cause? Note also that the
case where both the slow and fast subsystems are classical
was also studied\cite{Gozzi1987PRD} and a vector potential
related to Hannay's angle\cite{Hannay1985JPA} was found.

In this Letter we present a general theoretical framework
for the hybrid systems.  For this kind of system,
we use a well-known fact\cite{Heslot1985PRD,Weinberg1989AP,Liu2003PRL} 
that a quantum system possesses mathematically a canonical classical
Hamiltonian structure. In this way, we can describe the hybrid
system with a unified classical Hamiltonian. We emphasize that
in our Hamiltonian the quantum subsystem is reduced
only mathematically to a classical system and no physics is lost.
This is in contrast to the Hamiltonian of Ref.\cite{Berman2002PRA},
where the quantum subsystem is reduced physically to be 
classical and some physics may be lost.

We find that a vector potential can arise
in the hybrid system when the quantum subsystem is fast
and the classical subsystem is slow. This is achieved with 
a simple canonical transformation. On one hand, this vector
potential enters the fast quantum subsystem in terms of  the
familiar Berry phase\cite{Berry1984PRS}; on the other hand, it
produces a Lorentz-like force in the slow classical subsystem.
This implies that the phase of a wavefunction can lead to
direct observable phenomenon via a physical force,
augmenting significantly the usual understanding  that 
the phase can lead to direct
physical consequences only via interference.

We shall use a simple example, the coupling of a heavy magnetic
particle with a single spin, to illustrate our theory. In this example,
one can see clearly that the phase of a wavefunction, a non-real space variable,
can break the left-right symmetry in the real space.
The possibility of detecting such a force is also examined.

In addition, we point out that our method can also be regarded
as an alternative way of deriving the Berry phase\cite{Berry1984PRS}.
Our method has an advantage that the fast quantum subsystem needs not
be in an eigenstate. In other words, the Berry phase
can be defined for a general quantum state. 
Our method can also be generalized to derive the Hannay's angles\cite{Hannay1985JPA} 
or the geometric phase proposed in Ref.\cite{Wu2005PRL} for
nonlinear quantum systems.

The aforementioned hybrid coupled system can be described by the following
Hamiltonian,
\begin{equation}
H=\langle\bm{\Psi}|\hat{H}_1(\bm{q}_{2})|\bm{\Psi}\rangle+
H_{2}\left(\bm{p}_{2},\bm{q}_{2}
\right),
\end{equation}
where $\hat{H}_1$ is the Hamiltonian operator of the linear
$N$-level fast quantum subsystem and
$|\bm{\Psi}\rangle=(\Psi_{1},\Psi_{2},\cdots,\Psi_{N})^{T}$ is its
quantum state.  The Hamiltonian $H_{2}$ governs a heavy classical
subsystem that moves slowly and $\bm{p}_{2}$, $\bm{q}_{2}$ are its
momenta and coordinates, respectively. The dependence of $\hat{H}_1$
on $\bm{q}_{2}$ indicates the coupling between the two subsystems.

We first focus on the quantum subsystem assuming temporarily that
$\bm{q}_{2}$ are just some fixed parameters.
The Schr\"odinger equation of the quantum subsystem is
$i\hbar d|\bm{\Psi}\rangle/dt=\hat{H}_1|\bm{\Psi}\rangle$,
which can be re-written as
\begin{equation}
\label{a}
i\hbar\frac{d\Psi_{j}}{dt}=\frac{\partial}{\partial\Psi_{j}^{*}}H_{1}
\left(\bm{\Psi},\bm{\Psi}^{*},\bm{q}_{2}\right)\,,
\end{equation}
where $H_{1}=\langle\bm{\Psi}|\hat{H}(\bm{q}_{2})|\bm{\Psi}\rangle$.
This shows that the quantum system has a classical Hamiltonian structure.
This fact is known to many people and was discussed in details
in Ref.\cite{Heslot1985PRD,Weinberg1989AP}. To have a more ``classical''
look for this quantum system,  we define
$p_{1j}=\sqrt{i\hbar}\Psi_{j}^{*}$, $q_{1j}=\sqrt{i\hbar}\Psi_{j}$
and write Eq.(\ref{a}) in an apparent
canonical Hamiltonian formalism
\begin{equation}
\frac{dq_{1j}}{dt}=\frac{\partial \widetilde{H}_{1} }{\partial p_{1j}}\,,
\hspace{0.5cm}
\frac{dp_{1j}}{dt}=-\frac{\partial \widetilde{H}_{1}}{\partial q_{1j}}\,,
\end{equation}
where $\widetilde{H}_1=\widetilde{H}_1(\bm{p}_{1},\bm{q}_{1},\bm{q}_{2})=
H_{1}(\bm{\Psi},\bm{\Psi}^{*},\bm{q}_{2})$. In this way,
we have classically reformulated quantum systems.

The quantum state $|\bm{\Psi}\rangle$ can be expanded in terms of
instantaneous eigenstates 
\be
|\bm{\Psi}\rangle=\sum_{n=1}^{N}a_n|\varphi_n(\bm{q}_2)\rangle \ee
where $\hat{H}_1(\bm{q}_2)|\varphi_n(\bm{q}_2)\rangle=
E_n(\bm{q}_2)|\varphi_n(\bm{q}_2)\rangle$. The quantum system can be
described alternatively by these expansion coefficiencies $a_n$'s.
Define $I_{1n}=\hbar |a_n|^2$ and $\Theta_{1n}=-\arg(a_n)$; one can
prove readily that $\bm{I}_{1}$ and $\bm{\Theta}_{1}$ are another
set of canonical variables for the Hamiltonian $H_1$. According to
the standard classical theory\cite{Arnold1978Book}, there is a
canonical transformation between $\bm{p}_{1}$, $\bm{q}_{1}$ and
$\bm{I}_{1}$, $\bm{\Theta}_{1}$, and this transformation is given by
a generating function $F_{1}(\bm{q}_{1},\bm{I}_{1},\bm{q}_{2})$
that satisfies
\be
\label{eq:f1}
\bm{p}_{1}=\frac{\partial F_{1}}{\partial
\bm{q}_{1}}\,, \hspace{0.5cm} \bm{\Theta}_{1}=\frac{\partial
F_{1}}{\partial \bm{I}_{1}}\,.
\ee
With this transformation,
the Hamiltonian $H_1$ becomes ${\mathcal H}_1={\mathcal
H}_1(\bm{I}_{1},\bm{q}_2)= \sum_n E_n(\bm{q}_2)I_{1n}/\hbar$,
independent of the angles $\bm{\Theta_1}$.

We go back to the hybrid coupled system, where $\bm{q}_2$ are
dynamical variables instead of some fixed parameters. The Hamiltonian for
the hybrid system can be now expressed in a pure classical formalism
\begin{equation}
\label{eq:tham}
\widetilde{H}=\widetilde{H}_{1}\left(\bm{p}_{1},\bm{q}_{1},\bm{q}_{2}
\right)+H_{2}\left(\bm{p}_{2},\bm{q}_{2} \right)\,.
\end{equation}
We introduce a canonical transformation from $\bm{p}_{1}$, $\bm{q}_{1}$,
$\bm{p}_{2}$, $\bm{q}_{2}$ to $\bm{I}_{1}$, $\bm{\Theta}_{1}$,
$\bm{P}_{2}$, $\bm{Q}_{2}$ with the following
generating function
\be
F=F_{1}(\bm{q}_{1},\bm{I}_{1},\bm{q}_{2})+\bm{q}_{2}\bm{P}_{2}.
\ee
The canonical transformation is then given by
\be
\bm{p}_{1}=\frac{\partial F}{\partial \bm{q}_{1}}=
\frac{\partial F_{1}}{\partial \bm{q}_{1}}\,,~~~~
\bm{\Theta}_{1}=\frac{\partial F}{\partial \bm{I}_{1}}=\frac{\partial
F_{1}}{\partial \bm{I}_{1}}\,,
\label{eq:f12}
\ee
\be
\bm{p}_{2}=\frac{\partial F}{\partial \bm{q}_{2}}=\frac{\partial
F_{1}}{\partial \bm{q}_{2}}+\bm{P}_{2}\,,~~~~~
\bm{Q}_{2}=\frac{\partial F}{\partial \bm{P}_{2}}=\bm{q}_{2}\,.
\ee
The transformation does two things: (1) Since Eq.(\ref{eq:f12})
is identical to Eq.(\ref{eq:f1}), the transformation changes
$\bm{p}_{1}$, $\bm{q}_{1}$ to $\bm{I}_{1}$, $\bm{\Theta}_{1}$
as if it is generated by $F_1$; (2) it puts an additional
vector function $\bm{A}=-\partial F_{1}/\partial \bm{q}_{2}$
in the momenta $\bm{p}_2$ while keeping the coordinates
$\bm{q}_2$ unchanged. After the transformation, 
the total Hamiltonian in Eq.(\ref{eq:tham}) becomes
\begin{equation}
{\mathcal H}={\mathcal H}_{1}(\bm{I}_{1},\bm{Q}_{2})+
H_{2}(\bm{P}_{2}-\bm{A},\bm{Q}_{2})\,.
\end{equation}
In the second subsystem, $\bm{A}$ appears very much like  a vector
potential. However, it is not a true vector potential since it also 
depends on variables other than $\bm{q}_2$. For convenience, we shall call it
pseudo-vector potential. As we shall see, this pseudo-vector
potential can lead to a true vector potential.

By assuming that the classical subsystem has
the usual Hamiltonian $H_2=\bm{p}_2^2/2M+V_2(\bm{q}_2)$,
we write down the equations of motion for the whole system.
For the quantum part, we have
\begin{eqnarray}
\label{b} \dot{I}_{1j}&=&\dot{\bm{q}}_{2}\cdot\frac{\partial
\bm{A}}{\partial
\Theta_{1j}}\,, \\
\dot{\Theta}_{1j}&=&\frac{\partial {\mathcal H}_{1}}{\partial
I_{1j}}-\dot{\bm{q}}_{2}\cdot \frac{\partial \bm{A}}{\partial
I_{1j}}\,. \label{eq:qph}
\end{eqnarray}
For the classical part, we obtain
\begin{eqnarray}
\label{eq:cp} \dot{P}_{2j}&=&-\frac{\partial {\mathcal
H}_{1}}{\partial q_{2j}}-\frac{\partial V_{2}}{\partial
q_{2j}}+\dot{\bm{q}}_{2}\cdot\frac{\partial \bm{A}}{\partial
q_{2j}}\,, \\\label{eq:cp2}
\dot{Q}_{2j}&=&\dot{q}_{2j}=(P_{2j}-A_j)/M\,.
\end{eqnarray}
As we shall see, the last term in Eq.(\ref{eq:qph}) gives rise
to the familiar Berry phase while Eq.(\ref{eq:cp}) contains a
Lorentz-like force generated by the vector potential.

To analyze the pseudo-vector potential $\bm{A}$, we first make the
potential an explicit function of
the new dynamical variables
$\bm{I}_{1}$, $\bm{\Theta}_{1}$, $\bm{P}_{2}$, and $\bm{Q}_{2}(=\bm{q_2})$
by defining a new function,
\begin{equation}
\widetilde{F_{1}}(\bm{I}_{1},\bm{\Theta}_{1},\bm{q}_{2})=
F_{1}(\bm{q}_{1}(\bm{I}_{1},\bm{\Theta}_{1},\bm{q}_{2}),
\bm{I}_{1},\bm{q}_{2})\,.
\end{equation}
This leads to
\begin{equation}
A_{j}=\bm{p}_{1}\cdot\frac{\partial \bm{q}_{1}}{\partial
q_{2j}}-\frac{\partial \widetilde{F_{1}}}{\partial
q_{2j}}
=i\hbar\langle\bm{\Psi}|\frac{\partial}{\partial
q_{2j}}|\bm{\Psi}\rangle-\frac{\partial \widetilde{F_{1}}}{\partial
q_{2j}}\,,
\end{equation}
where we already see the indication of the Berry phase
if $|\bm{\Psi}\rangle$ is an eigenstate.

As we have assumed from the beginning that the quantum subsystem is fast
and the classical subsystem is slow, the variables $\bm{q}_2$ can
be regarded as adiabatic parameters for the quantum subsystem.
As a result, we are allowed to apply the standard averaging
technique in the study of adiabatic
evolution\cite{Hannay1985JPA,Arnold1978Book}.
It also implies that the probabilities ${\bm I}_1$ are
conserved during the evolution according to the quantum
adiabatic theorem\cite{Messiah1958Book}.
After the averaging and using the quantum adiabatic theorem,
the pseudo-vector potential becomes
\begin{eqnarray}
\overline{A}_{j}&=&\oint\frac{d\bm{\Theta}_{1}}{(2\pi)^{N}}
\Big[i\hbar\langle\bm{\Psi}|\frac{\partial}{\partial
q_{2j}}|\bm{\Psi}\rangle-\frac{\partial \widetilde{F_{1}}}{\partial
q_{2j}}
\Big]\,,\nonumber \\
&=&i\sum_{n=1}^N I_{1n}\langle\varphi_{n}|\frac{\partial}{\partial
q_{2j}}|\varphi_{n}\rangle- \overline{\frac{\partial
\widetilde{F_{1}}}{\partial q_{2j}}}\,, \label{eq:aaa}
\end{eqnarray}
where the overline indicates that the average has been done for
the variable. The function $\overline{\bm A}$ is now a true
vector potential as it no longer depends on $\bm{\Theta}_1$
and at the same time ${\bm I}_1$ are constant. After
ignoring the trivial gradient term  in Eq.(\ref{eq:aaa})
we obtain a true vector potential
\be
\overline{\bm A}=\sum_{n=1}^{N}I_{1n}{\bm A}_n\,,~~~
{\bm A}_n=i\langle\varphi_{n}|\frac{\partial}{\partial
{\bm q}_{2}}|\varphi_{n}\rangle\,,
\ee
Substituting it into Eq.(\ref{eq:qph}), we arrive at
\begin{equation}
\dot{\Theta}_{1j}=\frac{\partial {\mathcal H}_{1}}{\partial I_{1j}}-
{\bm A}_j\cdot\dot{\bm{q}}_{2}\,,
\end{equation}
where the integration of the last term produces exactly
the Berry phase of the $j$th eigenstate. One can
regard this as a new way to derive the Berry phase;
in this new way, the quantum system does not need to be
in an eigenstate.

For the slow classical subsystem, we take the averaging
over Eq.(\ref{eq:cp}) and rewrite it in a physically
more transparent form
\begin{equation}
\label{eq:yun}
M\overline{\ddot{\bm{q}}_{2}}=-\frac{\partial {\mathcal H}_{1}}
{\partial\bm{q}_{2}}-\frac{\partial V_{2}}{\partial
\bm{q}_{2}}+\overline{\dot{\bm{q}}}_{2}\times \bm{\mathcal B},
\end{equation}
where $\bm{\mathcal B}=\nabla\times \overline{\bm{A}}
=\sum_{n}I_{1n}\nabla\times{\bm A}_n$ is a
magnetic-like gauge field. Similar to the
usual magnetic field, the gauge field influences the
dynamics in terms of a Lorentz force. So, the
Berry phase is shown to be linked to a physical force.
We shall have more discussion with this force later
via an example.

\setlength{\unitlength}{1mm}
\begin{figure}[!tb]
\begin{picture}(60,40)
\put(5,5){\vector(1,0){10}} \put(5,5){\vector(0,1){10}}
\put(5,5){\vector(-1,-1){5}} \put(15,4){\makebox(0,0)[t]{$y$}}
\put(6,15){\makebox(0,0)[l]{$z$}} \put(1,0){\makebox(0,0)[l]{$x$}}
\thicklines \put(25,35){\line(1,0){45}} \put(15,15){\line(1,0){45}}
\put(25,35){\line(-1,-2){10}} \put(70,35){\line(-1,-2){10}}
\qbezier[1000](30,25)(30,30)(45,30)
\qbezier[1000](60,25)(60,30)(45,30)
\qbezier[1000](30,25)(30,20)(45,20)
\qbezier[1000](60,25)(60,20)(45,20) \thicklines
\put(45,25){\vector(1,0){15}}
\put(53,26){\makebox(0,0)[b]{$r$}}\put(45,27){\makebox(0,0)[tr]{$O$}}
\put(40,34){\makebox(0,0)[tr]{$xy$ plane}}
\put(34,21.2){\circle*{2}}
\put(29,20){\makebox(0,0)[t]{$\bm{m}_{F}$}}
\put(34,22){\vector(0,-1){5}} \thinlines
\multiput(45,25)(0,-3){7}{\line(0,-1){2}}\thicklines
\put(45,4){\circle*{1}} \put(42,1){\vector(1,1){6}}
\put(48,4){\makebox(0,0)[l]{$\bm{\mu}$(spin)}}
\put(46,13){\makebox(0,0)[l]{$d$}}
\end{picture}
\caption{A schematic setup of a magnetic particle interacting with a
spin. The particle moves freely in the $xy$ plane with a magnetic
moment of $\bm{m}_{F}$ always pointing in the negative $z$
direction. A single spin with magnetic moment $\bm{\mu}$ is placed
beneath the plane with a distance of $d$.}
\label{fig:bs}
\end{figure}
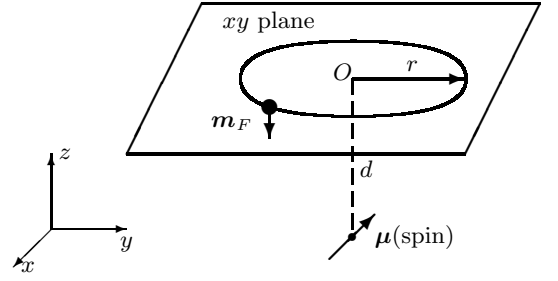

We have applied our method to the cantilever-spin system
in the single spin detection experiment\cite{Rugar2004Nature,Mamin2003PRL}
and recovered the theoretical result in Ref.\cite{Berman2002PRA}.
The pity is that since the cantilever vibrates only in one dimension,
the force associated with $\overline{\bm A}$ is always zero.
To have a nonzero $\bm{\mathcal B}$, we
consider an example, where the classical subsystem is a magnetic
particle and the quantum subsystem is a spin of 1/2 as shown in
Fig.\ref{fig:bs}. The magnetic particle has a magnetic moment
$\bm{m}_{F}$ and mass $m$; it moves freely in the $xy$ plane. We
assume that the magnetic moment $\bm{m}_{F}$ always points in the
negative $z$ direction. A single spin with magnetic moment
$\bm{\mu}$ is placed below the plane at the distance of $d$. For
simplicity, we place the origin of our coordinate system in the
particle plane and directly above the spin.

Due to the magnetic dipolar interaction, the spin feels a magnetic
field from the classical magnetic particle. The field is given by
\be
\{B_x,B_y,B_z\}=-\frac{\mu_{0}m_{F}\{3xd\,,3yd\,,2d^2-r^2\}}
{4\pi(d^{2}+r^{2})^{\frac{5}{2}}}\,,
\ee
where $r^2=x^2+y^2$. So, the Hamiltonian operator for the spin is
\begin{equation}
\hat{H}_1=-\mu\left(\begin{array}{cc}B_{z}&B_{x}-iB_{y}\\
B_{x}+iB_{y}&-B_{z}\end{array}\right).
\end{equation}
This Hamiltonian has two eigenstates $|\pm\rangle$, whose
eigen-energies are respectively $\mp \mu B$ with
$B=\sqrt{B_x^2+B_y^2+B_z^2}$.
The total Hamiltonian is
\be
H=\langle\bm{\Psi}|\hat{H}_1|\bm{\Psi}\rangle+\bm{p}^2/2m\,,
\ee
where $|\bm{\Psi}\rangle=(\Psi_1,\Psi_2)^T$ is the spin wave
function and $\bm{p}$ is the momentum of the magnetic particle.

Following the general procedure described above, we can transform
the above Hamiltonian to \be {\mathcal H}=(|a_{-}|^2-|a_{+}|^2)\mu
B+ \frac{\left(\bm{P}-\overline{\bm{A}}\right)^{2}}{2m}\,, \ee where
$|a_{\pm}|^2$ are the probabilities on the two spin eigenstates
$|\pm\rangle$, respectively. The vector potential is given by
\begin{equation}
\overline{\bm{A}}=i\hbar|a_{+}|^2\langle +|\frac{\partial}{\partial
\bm{r}}|+\rangle +i\hbar|a_{-}|^2\langle -|\frac{\partial}{\partial
\bm{r}}|-\rangle\,.
\end{equation}
The equations of motion for the particle are
\begin{eqnarray}
\label{eq:yundong1}
m\ddot{x}=\frac{3\mu\mu_{0}m_{F}(5d^{2}+r^{2})(|a_{-}|^2-|a_{+}|^2)}
{4\pi\sqrt{4d^{2}+r^{2}}(d^{2}+r^{2})^{3}}x+{\mathcal B}\dot{y}\,,\\
\label{eq:yundong2}
m\ddot{y}=\frac{3\mu\mu_{0}m_{F}(5d^{2}+r^{2})(|a_{-}|^2-|a_{+}|^2)}
{4\pi\sqrt{4d^{2}+r^{2}}(d^{2}+r^{2})^{3}}y-{\mathcal B}\dot{x}\,.
\end{eqnarray}
The magnetic-like $\bm{\mathcal B}$ field always points in the $z$ direction.
The field strength is
\begin{equation}
{\mathcal B}=\frac{9\hbar d^{2}(r^{2}+2d^{2})}
{2\left[(r^{2}+d^{2})(r^{2}+4d^{2})\right]^{3/2}}(|a_{+}|^2-|a_{-}|^2).
\end{equation}

We make two observations. First, the field is geometric,
depending only on the position of the magnetic particle besides
$\hbar$ and independent of the strength of the dipolar
interaction. Secondly, it curves motion in an unexpected way. For
instance, if the initial conditions of the slow particle are
$x(0)=0$, $y(0)=0$, $\dot{x}(0)>0$, and $\dot{y}(0)=0$, then
everything in the real space is symmetric with respect to the
inverse of $y$, including the dipolar force and the spin direction.
One would then expect intuitively the particle move in a straight
line along the $x$-direction. However, the slow particle will curve
owing to $\mathcal B$, breaking the left-right symmetry in the
real space. What is interesting is that the force breaking of this 
real-space symmetry comes from the dynamics of the 
spin wavefunction's phase, a non-real space variable.

To detect such a force experimentally, the best way
may be to measure the frequency associated with this force, such as the
frequency change in the single spin detection\cite{Rugar2004Nature}.
For the simple system in Fig.\ref{fig:bs}, we notice that the slow
particle can move in a circle if $\mu<0$. If somehow one can fix the
radius $r$ of the circle, the frequency that the particle circulates
clockwise is different from anticlockwise. Simple calculations show
that the frequency difference is $\Delta\nu={\mathcal B}/2\pi m$.
For estimation of the value of the frequency difference, the
following parameters are used:
$\mu_{0}m_{F}\sim2.0\times10^{-21}$T$\cdot$m$^{3}$,
$m\sim2.5\times10^{-15}$kg, $r\sim1$nm, and $d\sim1000$nm. For these
values of parameters, we obtain ${\mathcal
B}\sim-1.20\times10^{-22}$kg/s (if it were for an electron, it is
equivalent to a magnetic field of
 $7.5\times10^{-4}$T) and $Bz\sim3.2\times10^{-4}$T.
The frequency difference is $\Delta\nu\sim0.7\times10^{-8}$Hz, which
is a challenging task for the current
technique\cite{Rugar2004Nature}. Moreover, in a real experiment, the
magnetic particle needs to be attached to something such as a string
so that it can oscillate in two dimensions.

A similar vector potential can also arise in connection to Hannay's
angle\cite{Hannay1985JPA}. The theoretical framework starting at 
Eq.(\ref{eq:tham}) can be readily
generalized to the cases where the fast subsystem is a
classical integrable system.
In this generalization, one only needs to regard $\bm{I}_1$ and
$\bm{\Theta}_1$ as a set of action-angle variables.
Such a coupled system was also discussed in Ref.\cite{Gozzi1987PRD}.
The same extension can be done for the geometric
phase proposed for nonlinear quantum systems\cite{Wu2005PRL}.

We point out that our method with the canonical transformation
can also be regarded as a unified theoretical framework for deriving
Berry phase\cite{Berry1984PRS}, Hannay's angle\cite{Hannay1985JPA}, the new
geometric phase for nonlinear quantum systems\cite{Wu2005PRL}.
We further note that, as one may have noticed, 
even in the general case where the two subsystems are not
necessarily one fast and one slow, one may
use the pseudo-vector potential $\bm{A}$ to define a ``geometric'' phase. 
Such defined phase is obviously for non-adiabatic
processes but different from the Aharonov-Anandan phase\cite{AA}.
The important point is that this general phase may be of little
use since it does not provide any insight into the dynamics.

It is worthwhile to note that a similar force also appears
in the semiclassical dynamics of Bloch
electrons\cite{Chang1995PRL,Sundaram1999PRB}. However,
in this system, both the fast and the slow dynamics are
for Bloch electrons, there are no different subsystems.
With all these results, one cannot help but wonder whether
the vector potential in electrodynamics also has a dynamical origin.
If so, this may explain why there is no magnetic charge.

We thank Junren Shi, Di Xiao, and Qian Niu for helpful discussion.
This work is supported by the ``BaiRen'' program of the CAS, 
NSF of China(10504040), and the
973 project (2005CB724508).


\begin{thebibliography}{21}
\expandafter\ifx\csname natexlab\endcsname\relax\def\natexlab#1{#1}\fi
\expandafter\ifx\csname bibnamefont\endcsname\relax
  \def\bibnamefont#1{#1}\fi
\expandafter\ifx\csname bibfnamefont\endcsname\relax
  \def\bibfnamefont#1{#1}\fi
\expandafter\ifx\csname citenamefont\endcsname\relax
  \def\citenamefont#1{#1}\fi
\expandafter\ifx\csname url\endcsname\relax
  \def\url#1{\texttt{#1}}\fi
\expandafter\ifx\csname urlprefix\endcsname\relax\def\urlprefix{URL }\fi
\providecommand{\bibinfo}[2]{#2}
\providecommand{\eprint}[2][]{\url{#2}}

\bibitem[{\citenamefont{Rugar et~al.}(2004)\citenamefont{Rugar, Budakian,
  Mamin, and Chui}}]{Rugar2004Nature}
\bibinfo{author}{\bibfnamefont{D.}~\bibnamefont{Rugar}},
  \bibinfo{author}{\bibfnamefont{R.}~\bibnamefont{Budakian}},
  \bibinfo{author}{\bibfnamefont{H.~J.} \bibnamefont{Mamin}}, \bibnamefont{and}
  \bibinfo{author}{\bibfnamefont{B.~W.} \bibnamefont{Chui}},
  \bibinfo{journal}{Nature} \textbf{\bibinfo{volume}{430}},
  \bibinfo{pages}{329} (\bibinfo{year}{2004}).

\bibitem[{\citenamefont{Hammel}(2004)}]{Hammel2004Nature}
\bibinfo{author}{\bibfnamefont{P.~C.} \bibnamefont{Hammel}},
  \bibinfo{journal}{Nature} \textbf{\bibinfo{volume}{430}},
  \bibinfo{pages}{300} (\bibinfo{year}{2004}).

\bibitem[{\citenamefont{Slichter}(1990)}]{SlichterBook}
\bibinfo{author}{\bibfnamefont{C.~P.} \bibnamefont{Slichter}},
  \emph{\bibinfo{title}{Principles of Magnetic Resonance}}
  (\bibinfo{publisher}{Springer, Berlin}, \bibinfo{year}{1990}).

\bibitem[{\citenamefont{Berman et~al.}(2002)\citenamefont{Berman, Kamenev, and
  Tsifrinovich}}]{Berman2002PRA}
\bibinfo{author}{\bibfnamefont{G.~P.} \bibnamefont{Berman}},
  \bibinfo{author}{\bibfnamefont{D.~I.} \bibnamefont{Kamenev}},
  \bibnamefont{and} \bibinfo{author}{\bibfnamefont{V.~I.}
  \bibnamefont{Tsifrinovich}}, \bibinfo{journal}{Phys. Rev. A}
  \textbf{\bibinfo{volume}{66}}, \bibinfo{pages}{023405}
  (\bibinfo{year}{2002}).

\bibitem[{\citenamefont{Mamin et~al.}(2003)\citenamefont{Mamin, Budakian, Chui,
  and Rugar}}]{Mamin2003PRL}
\bibinfo{author}{\bibfnamefont{H.~J.} \bibnamefont{Mamin}},
  \bibinfo{author}{\bibfnamefont{R.}~\bibnamefont{Budakian}},
  \bibinfo{author}{\bibfnamefont{B.~W.} \bibnamefont{Chui}}, \bibnamefont{and}
  \bibinfo{author}{\bibfnamefont{D.}~\bibnamefont{Rugar}},
  \bibinfo{journal}{Phys. Rev. Lett.} \textbf{\bibinfo{volume}{91}},
  \bibinfo{pages}{207604} (\bibinfo{year}{2003}).

\bibitem[{\citenamefont{Mead and Truhlar}(1979)}]{Mead1979JCP}
\bibinfo{author}{\bibfnamefont{C.~A.} \bibnamefont{Mead}} \bibnamefont{and}
  \bibinfo{author}{\bibfnamefont{D.~G.} \bibnamefont{Truhlar}},
  \bibinfo{journal}{J. Chem. Phys.} \textbf{\bibinfo{volume}{70}},
  \bibinfo{pages}{2284} (\bibinfo{year}{1979}).

\bibitem[{\citenamefont{Aharonov et~al.}(1990)\citenamefont{Aharonov,
  Ben-Reuven, Popescu, and Rohrlich}}]{Aharonov1990PRL}
\bibinfo{author}{\bibfnamefont{Y.}~\bibnamefont{Aharonov}},
  \bibinfo{author}{\bibfnamefont{E.}~\bibnamefont{Ben-Reuven}},
  \bibinfo{author}{\bibfnamefont{S.}~\bibnamefont{Popescu}}, \bibnamefont{and}
  \bibinfo{author}{\bibfnamefont{D.}~\bibnamefont{Rohrlich}},
  \bibinfo{journal}{Phys. Rev. Letts.} \textbf{\bibinfo{volume}{65}},
  \bibinfo{pages}{3065} (\bibinfo{year}{1990}).

\bibitem[{\citenamefont{Kuratsuji and Iida}(1985)}]{Iida1985PTP}
\bibinfo{author}{\bibfnamefont{H.}~\bibnamefont{Kuratsuji}} \bibnamefont{and}
  \bibinfo{author}{\bibfnamefont{S.}~\bibnamefont{Iida}},
  \bibinfo{journal}{Prog. Theor. Phys.} \textbf{\bibinfo{volume}{74}},
  \bibinfo{pages}{439} (\bibinfo{year}{1985}).

\bibitem[{\citenamefont{Stone}(1986)}]{Stone1986PRD}
\bibinfo{author}{\bibfnamefont{M.}~\bibnamefont{Stone}},
  \bibinfo{journal}{Phys. Rev. D} \textbf{\bibinfo{volume}{33}},
  \bibinfo{pages}{1191} (\bibinfo{year}{1986}).

\bibitem[{\citenamefont{Berry}(1984)}]{Berry1984PRS}
\bibinfo{author}{\bibfnamefont{M.~V.} \bibnamefont{Berry}},
  \bibinfo{journal}{Proc. R. Soc. Lond. A} \textbf{\bibinfo{volume}{392}},
  \bibinfo{pages}{45} (\bibinfo{year}{1984}).

\bibitem[{\citenamefont{Gozzi and Thacker}(1987)}]{Gozzi1987PRD}
\bibinfo{author}{\bibfnamefont{E.}~\bibnamefont{Gozzi}} \bibnamefont{and}
  \bibinfo{author}{\bibfnamefont{W.~D.} \bibnamefont{Thacker}},
  \bibinfo{journal}{Phys. Rev. D} \textbf{\bibinfo{volume}{35}},
  \bibinfo{pages}{2398} (\bibinfo{year}{1987}).

\bibitem[{\citenamefont{Hannay}(1985)}]{Hannay1985JPA}
\bibinfo{author}{\bibfnamefont{J.~H.} \bibnamefont{Hannay}},
  \bibinfo{journal}{J. Phys. A} \textbf{\bibinfo{volume}{18}},
  \bibinfo{pages}{221} (\bibinfo{year}{1985}).

\bibitem[{\citenamefont{Heslot}(1985)}]{Heslot1985PRD}
\bibinfo{author}{\bibfnamefont{A.}~\bibnamefont{Heslot}},
  \bibinfo{journal}{Phys. Rev. D} \textbf{\bibinfo{volume}{31}},
  \bibinfo{pages}{1341} (\bibinfo{year}{1985}).

\bibitem[{\citenamefont{Weinberg}(1989)}]{Weinberg1989AP}
\bibinfo{author}{\bibfnamefont{S.}~\bibnamefont{Weinberg}},
  \bibinfo{journal}{Ann. Phys. (N.Y.)} \textbf{\bibinfo{volume}{194}},
  \bibinfo{pages}{336} (\bibinfo{year}{1989}).

\bibitem[{\citenamefont{Liu et~al.}(2003)\citenamefont{Liu, Wu, and
  Niu}}]{Liu2003PRL}
\bibinfo{author}{\bibfnamefont{J.}~\bibnamefont{Liu}},
  \bibinfo{author}{\bibfnamefont{B.}~\bibnamefont{Wu}}, \bibnamefont{and}
  \bibinfo{author}{\bibfnamefont{Q.}~\bibnamefont{Niu}},
  \bibinfo{journal}{Phys. Rev. Lett.} \textbf{\bibinfo{volume}{90}},
  \bibinfo{pages}{170404} (\bibinfo{year}{2003}).

\bibitem[{\citenamefont{Wu et~al.}(2005)\citenamefont{Wu, Liu, and
  Niu}}]{Wu2005PRL}
\bibinfo{author}{\bibfnamefont{B.}~\bibnamefont{Wu}},
  \bibinfo{author}{\bibfnamefont{J.}~\bibnamefont{Liu}}, \bibnamefont{and}
  \bibinfo{author}{\bibfnamefont{Q.}~\bibnamefont{Niu}},
  \bibinfo{journal}{Phys. Rev. Lett.} \textbf{\bibinfo{volume}{94}},
  \bibinfo{pages}{140402} (\bibinfo{year}{2005}).

\bibitem[{\citenamefont{Arnold}(1978)}]{Arnold1978Book}
\bibinfo{author}{\bibfnamefont{V.~I.} \bibnamefont{Arnold}},
  \emph{\bibinfo{title}{Mathematical Methods of Classical Mechanics}}
  (\bibinfo{publisher}{Springer-Verlag}, \bibinfo{year}{1978}).

\bibitem[{\citenamefont{Messiah}(1958)}]{Messiah1958Book}
\bibinfo{author}{\bibfnamefont{A.}~\bibnamefont{Messiah}},
  \emph{\bibinfo{title}{Quantum Mechanics}} (\bibinfo{publisher}{Dover, New
  York}, \bibinfo{year}{1958}).

\bibitem[{\citenamefont{Aharonov and Anandan}(1987)}]{AA}
\bibinfo{author}{\bibfnamefont{Y.}~\bibnamefont{Aharonov}} \bibnamefont{and}
  \bibinfo{author}{\bibfnamefont{J.}~\bibnamefont{Anandan}},
  \bibinfo{journal}{Phys. Rev. Lett.} \textbf{\bibinfo{volume}{58}},
  \bibinfo{pages}{1593} (\bibinfo{year}{1987}).

\bibitem[{\citenamefont{Chang and Niu}(1995)}]{Chang1995PRL}
\bibinfo{author}{\bibfnamefont{M.-C.} \bibnamefont{Chang}} \bibnamefont{and}
  \bibinfo{author}{\bibfnamefont{Q.}~\bibnamefont{Niu}},
  \bibinfo{journal}{Phys. Rev. Lett.} \textbf{\bibinfo{volume}{75}},
  \bibinfo{pages}{1348} (\bibinfo{year}{1995}).

\bibitem[{\citenamefont{Sundaram and Niu}(1999)}]{Sundaram1999PRB}
\bibinfo{author}{\bibfnamefont{G.}~\bibnamefont{Sundaram}} \bibnamefont{and}
  \bibinfo{author}{\bibfnamefont{Q.}~\bibnamefont{Niu}},
  \bibinfo{journal}{Phys. Rev. B} \textbf{\bibinfo{volume}{59}},
  \bibinfo{pages}{14915} (\bibinfo{year}{1999}).

\end{thebibliography}



\end{document}